\documentclass[useAMS,usenatbib]{mn2e}

\usepackage[dvips]{graphicx}
\usepackage{graphics}
\usepackage{psfig}
\usepackage{lscape}
\usepackage{amssymb}
\input{epsf}

\title[Evidence for relativistic features in the  X-ray spectrum of Mrk~335.]{Evidence for relativistic features in the X-ray spectrum of  Mrk~335.}
\author[A. L. Longinotti et al.]{A. L. Longinotti$^{1,2}$\thanks{E-mail:
anna.lia.longinotti@sciops.esa.int}, S.A. Sim$^{1,3}$, K. Nandra$^{1}$, M. Cappi$^{4}$
\\
$^{1}$Astrophysics Group, Imperial College London, Blackett Laboratory, Prince Consort Road, SW7 2AZ London, UK\\
$^{2}$ XMM-Newton Science Operation Centre, ESAC, ESA, Apartado 50727, E-28080 Madrid, Spain\\
$^{3}$ Max-Planck-Institut f{\"u}r Astrophysik, 85741 Garching, Germany\\
$^{4}$INAF-IASF Sezione di Bologna, VIa Gobetti 101, I-40129 Bologna, Italy\\
}
\begin{document}

\date{Accepted}

\pagerange{\pageref{firstpage}--\pageref{lastpage}} \pubyear{2005}

\maketitle

\label{firstpage}

\begin{abstract}
We present an  analysis of hard X-ray features in the spectrum of the  bright Sy~1 galaxy Mrk~335 observed by the {\it XMM-Newton} satellite. Our analysis confirms the presence of a broad, ionised
iron K$\alpha$ emission line in the spectrum, first found by Gondoin et al. The broad line can be modeled successfully by relativistic accretion disc reflection models. This interpretation is unusually robust  in the case of Mrk 335 because of the lack of any ionised (`` warm'') absorber and the absence a clear narrow core to the line. Partial covering by neutral gas cannot, however,  be ruled out statistically as the origin of the broad residuals. Regardless of the underlying continuum we report, for the first time in this source, the detection of a narrow absorption feature at the rest frame energy of $\sim$ 5.9 keV. If the feature is identified with a resonance absorption line of iron in a highly ionised medium, the redshift of the line corresponds to an  inflow velocity  of $\sim$ 0.11-0.15 {\it c}.  We present a simple model for the inflow, accounting approximately for relativistic and radiation pressure effects, and use Monte Carlo methods to compute synthetic spectra for qualitative comparison with the data. This modeling shows that the absorption feature can plausibly be reproduced by infalling gas providing that the feature is identified with Fe {\sc xxvi}. We require the inflowing gas to extend over a limited range of radii at a few tens of $r_{\rm g}$ to match the observed feature. The mass accretion rate in the flow corresponds to 60\% of the Eddington limit, in remarkable agreement with the observed rate. The narrowness of the absorption line tends to argue against a purely gravitational origin for the redshift of the line, but given the current data quality we stress that such an interpretation cannot be ruled out. 
\end{abstract}

\section{Introduction}
\label{sec:intro_335}
Since the discovery of Active Galactic Nuclei (AGN),  it has been postulated that the powering mechanism is likely to be the release of gravitational energy by matter accreted on a supermassive black hole (e.g. Lynden-Bell 1969). Evidence for material close to the black hole has been found in the redshifted and broad Fe K disc line seen in  bright Seyfert galaxies (Tanaka et al. 1995; Nandra et al. 1997).  An important concern for the disc line interpretation is that many Seyfert galaxies possess
highly ionised gas in the lines of sight (``warm absorbers'' e.g. Halpern 1984; George et al. 1998).
If there is an absorber  component  with sufficiently high ionisation state and column density, it can distort the continuum redward of the Fe K$\alpha$  line mimicking a broad line profile (e.g. Reeves et al. 2004).  Among the best examples of individual objects with broad lines are MCG-6-30-15 (Tanaka et al. 1995, Iwasawa et al. 1996, Wilms et al. 2001; Fabian et al. 2002, Ponti et al.  2004) and NGC 3516 (Nandra et al. 1999, Iwasawa et al. 2004). Both have ionised absorbers, but recent analysis of both objects have shown that despite their effects,  a broad line is still required in the data (Young et al. 2005, Turner et al. 2005).  If the remaining doubts about the effects of complex absorption can be dispelled, broad iron K$\alpha$ lines can be used as a diagnostic of the accretion flow in the strong gravity regime (e.g. Fabian et al. 2000; Reynolds \& Nowak 2003).  It would therefore be reassuring for the disc line interpretation if an object without a warm absorber could be found to exhibit broad iron emission. 

The broad iron lines represent evidence for rotation close to the central black hole in the form of an accretion disc. Observational evidence for inflow of matter in the vicinity of the  black hole remains extremely scarce, however. Nandra et al. (1999) reported the presence of a highly reshifted absorption feature identified with iron K$\alpha$ in the ASCA spectrum of NGC 3516. This was interpreted as being due to a high velocity inflow at $\sim 0.1c$, and thus representing evidence for matter flowing onto the black hole. An alternative interpretation in terms of gravitational redshift was presented by Ruszkowski \& Fabian (2000). Subsequent,  high signal-to-noise spectra of NGC 3516 have been presented by Turner et al. (2002), and showed narrow, shifted {\it emission} features, rather than absorption. However, the apparently transient nature of the narrow absorption and emission features (Nandra et al. 1999; Turner et al. 2002) make it hard to confirm or refute them conclusively. Very recently, a number of additional reports of redshifted absorption features have been made based on {\it BeppoSAX, XMM-Newton}  and {\it Chandra} data (Dadina et al. 2005; Matt et. al. 2005; Reeves et al. 2005; Yaqoob \& Serlemitsos 2005). The significance of these features is generally somewhat marginal, but the great increase in the number of reported cases since the initial detection in NGC 3516 increases the likelihood that  the redshifted absorption features are real and can be used as a meaningful tool to probe the accretion environment of the black hole.

 Mrk~335 is a bright Seyfert 1 galaxy at {\it z}=0.026, known also as PG 0003+199 as  part of the Palomar Green sample of UV-excess selected objects.
It was  observed several times by different X-ray observatories.
Observations by {\it EXOSAT} (Turner \& Pounds, 1989) and {\it  BBXRT} (Turner et al. 1993) reported the presence of a soft X-ray excess, which was confirmed later 
by Reynolds (1997) in a sample of sources observed by {\it ASCA}.
{\it ASCA} data did not provide any clear   evidence for warm absorption (Reynolds, 1997; George et al. 1998), but Turner et al. (1993b) found tentative evidence 
for soft X-ray spectral complexity in {\it ROSAT} data and Nandra \& Pounds (1994) found evidence for a hard X-ray edge  in {\it Ginga} data.

{\it BeppoSAX} data highlighted the presence of a reflection component 
and a  very strong Fe K line, interpreted as arising from an ionised accretion disc (Bianchi et al. 2001).
Ballantyne et al. (2001) have found that the {\it ASCA} spectrum of Mrk~335 is well fitted 
by an ionised disc model using the code by Ross \& Fabian (1993).

The {\it XMM-Newton} spectrum has been previously studied by  Gondoin et al. (2002)
who reported on   both EPIC and high resolution data from the RGS spectrometer in the 
0.3-2.1 keV band. From such work, the following findings have emerged:
the presence of  a relativistic Fe K line with an extremely high EW ($\sim$~800 eV), interpreted as originating from the inner regions of a disc in accretion around a  spinning  black hole; the low energy  X-ray continuum is characterised by a soft excess and it is interpreted as arising  from a combination of an  ionised reflection  and from intrinsic thermal emission in the disc; no evidence for soft X-ray features other than the 0.54 keV edge from Galactic Oxygen has been found in EPIC and in the RGS data. 

Here a re-analysis of Mrk~335 is presented.
The paper is organised as follows:  section 2 reports on the data reduction; section 3 describes the spectral analysis of the Fe K line and of the  X-ray continuum; section 4 reports on the detection of a narrow absorption line including a discussion on its significance; in section 5 a Monte Carlo model of
a relativistic inflow is described, for comparison with the observations. Results from the spectral fits and from the inflow model  are discussed in section 6.
\section{Observations and data reduction}
\label{sec:data_red_335}
Mrk~335 was observed by {\it XMM-Newton}  on 25-12-2000 (ID 0101040101) for a duration of $\sim$ 34 ks.
The data have been reduced with SAS version  (6.5.0) and calibrated event lists were obtained with the tasks epproc and emproc. The absence of soft protons flares in the light curve and  an extremely low background level  yield an exposure of $\sim$ 30 ks after accounting for the detector deadtime.
The observation was performed in full frame mode  for the pn camera and in large window mode for the MOS detectors. The high count rate of the source induced photon pile-up in all three
EPIC cameras. 
To reduce the effect of the pile-up in the spectra, the source counts were extracted  from annular regions in order to exclude the core of the PSF  where the most piled-up pixels are concentrated. 
The task epatplot  was employed to check the level of pile-up in the event files  obtained from  annular regions with increasing inner radius.
The pn  spectrum was extracted from an annular region with radii of  8 and 50 arcsec and the MOS spectra  were taken  from annuli 
with radii between 4 and 60 arcsec (i.e. a circle of  2 pixels radius  has been excluded). Single and double patterns were selected.
The background counts were chosen from a circular  source-free area close to the target  centroid, with extraction  radius of 50 and 60 arcsec.  The spectra  have been grouped in order to have at least 50 counts per bin in the pn and 20 counts per bin in the MOS spectra in order to apply $\chi^2$ minimization in the spectral fitting.

An inspection of the light curves shows that  the source flux varies by $\sim 20$ per cent 
over the whole exposure, but no strong spectral variability is detected.
The analysis is therefore carried out on the whole integrated spectrum.
Throughout the paper, the errors are quoted at 90\%  for one interesting parameter (i.e. $\Delta\chi^2$=2.71 criterion)
and the energies of spectral features are corrected for the redshift of the source.

\section{Spectral analysis of the broad feature  and the X-ray continuum}
{\label{sec:spec_an_335}
\begin{figure}
\psfig{figure=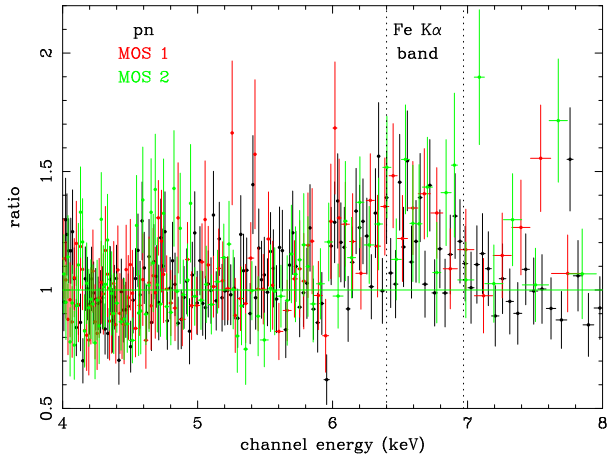,height=6cm,width=8cm}
\caption{\label{fig:ratio_335}Data to model ratio: the 2-10 keV spectrum is fitted by a power law with $\Gamma$= 2.15$^{+0.04}_{-0.03}$ . The plot shows the residuals of the EPIC spectra (black points: pn data; red points: MOS1; green points: MOS2) in the source rest frame and the Fe K$\alpha$ energy band is labelled for clarity.}
\end{figure}
We  begin by fitting the  pn and the MOS  spectra  between 
2-10~keV with the aim of finding a suitable  parametrization of the data.
The MOS1 and MOS2 spectra are fitted simultaneously.
 A simple power law  model yields a  steep spectrum, with the photon indeces 
 $\Gamma$=2.16$^{+0.02}_{-0.02}$  in the pn and  2.13$^{+0.03}_{-0.03}$ in the   MOS. 
 The $\chi^2$ probability values are marginally acceptable ($\chi^2$/d.o.f.=414/413 for the pn and $\chi^2$/d.o.f.=560/550 for the MOS, but the residuals from both instruments  
 show systematic deviations from the power law model. 
 The residuals  obtained from fitting  simultaneously  the EPIC data are plotted in  Fig. \ref{fig:ratio_335} (in the 4 to 8 keV band).
   The presence  of broad excess in flux 
 above the position of the neutral line (6.4 keV) up to  $\sim$ 7.2~keV  
 and down to $\sim$ 5.9 keV is quite clear, i.e. both ``blue" and ``red" flux are present in the residuals. There is no clear narrow core to the emission line, in constrast to several other similar Seyferts (Yaqoob \& Padmanhaban 2004). In addition to the excess emission component at the iron line,  a deficit of counts in a notch-shape is also  present at $\sim$ 5.9~keV.

A phenomenological description of the data is given in the following  before embarking on fitting more physical models.  The spectral fits are performed on the pn data only since the MOS sensitivity above 5 keV  is much lower than the pn. The MOS are nonetheless consistent with what is found in the pn. 
Firstly, a Gaussian line was added to the power law, with energy, width and flux free to vary.
 The line is  highly significant with $\Delta$$\chi^2$=38 for 3 d.o.f. 
 We find  a rest-frame energy of  E=6.49$^{+0.13}_{-0.31}$~keV, 
 $\sigma$=0.40$^{+0.65}_{-0.15}$ ~keV and an equivalent width of EW=~260$^{+240}_{-100}$~eV. 
Although the  residual shape in Fig.~\ref{fig:ratio_335} suggest the presence of a more complex line than this, any attempt to fit the data with 2 Gaussian emission lines failed to produce a statistically significant improvement.
 
To fit the notch-shaped absorption feature a Gaussian line with negative intensity  was added. The width of the absorption line is unresolved with CCD resolution and therefore  it is kept fixed to 1 eV. The line parameter in the pn camera  are  E=5.92$^{+0.04}_{-0.04}$~ keV with an EW=50$^{+18}_{-23}$~eV  (measured in absorption with negative intensity with respect to the continuum), and the broad line parameters change to E=6.22$^{+0.16}_{-0.18}$~keV,  $\sigma$=0.66$^{+0.27}_{-0.19}$~keV and EW=490$^{+230}_{-150}$~eV. The confidence contours for the absorption line are shown in Fig. \ref{fig:abs_line_335}. 
The improvement in $\chi^2$ is $\Delta\chi^2$$\sim$16 for 2 degrees
of freedom, corresponding to a level of confidence higher than 99.7 percent
according to the F-test.  
The fit yields an acceptable $\chi^2$=360/408 d.o.f. but it is clearly
only a basic  parametrization  of the spectrum. 
We have checked for the presence of the narrow line in the  MOS spectra, using the same baseline model as for the pn.
 The line parameters were found to be in agreement with the pn results: 
the energy is E=5.91$^{+0.12}_{-0.16}$~ keV and the EW= 32$^{+23}_{-28}$~eV, 
yielding therefore an upper limit consistent  with the value found in the pn.
The narrow line is less significant in the 
MOS ($\Delta\chi^2$$\sim$6) as expected giving that the collecting  area of the combined MOS is only $\sim$80 percent of the pn. 
We now go on to consider more physically motivated interpretations of the spectral features. 
\begin{figure}
\begin{center}
\psfig{figure=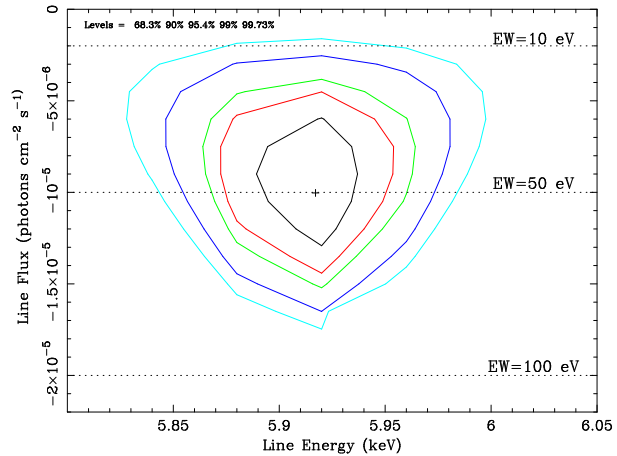,height=6cm,width=8cm}
\caption{\label{fig:abs_line_335} Confidence contours (68.3, 90, 95.4, 99 and 99.73 per cent) of  
the narrow absorption line (pn data); the intensity is measured from a  power law continuum with photon index $\Gamma$=2.16$\pm$0.04.
 The EW of the line corresponding to various values of photons flux is shown for clarity.} 
\end{center} 
\end{figure}

\subsection{A disc line origin for the Fe K line ?}
\label{subsec:diskline}
The line profile in Fig. \ref{fig:ratio_335} appears complex not only because of the 
absorption feature, but also because the residuals show a broad
and possible double-peaked structure. 
 The profile is in facts asymmetrical and skewed suggesting that it could be modified  by  relativistic effects. To test this hypothesis we have replaced the broad Gaussian with a  {\small DISKLINE} component (Fabian et al. 1989) and with  a {\small LAOR} component (Laor 1991). 
The former model corresponds to the  computation of the line photons 
emitted close to a non-spinning black hole in the Schwarzschild metric
whereas the latter adopts the Kerr metric, for a rapidly rotating black hole.   
For both models, the line parameters are the rest energy, $E$,
{\it q}, the line emissivity index,     
where the line emissivity {\it j} is a function of the emission radius {\it r}
according to  {\it j} $\propto${\it r$^{-q}$}; the inner radius {\it r$_{in}$}
and the outer radius  {\it r$_{out}$} of the accretion disc which define 
the area of the disc where the line is emitted; the inclination of the disc {\it i},  
defined as the angle between the line of sight and the normal to the disc. 

The two models were tested fixing the emissivity index {\it q}=3.
In both cases the spectral fits are consistent with the line being emitted at energy
higher than 6.4~keV, indicating an ionised accretion disc, confirming the results by Gondoin et al. 2002.
When the line profile is fitted with {\small DISKLINE}, we obtain  E=6.99$^{+0.70}_{-0.32}$~keV and EW=344$^{+90}_{-90}$~eV,  
{\it i}=21$^{\circ}$$^{+11}_{-13}$ and  $\chi^2$/d.o.f.=375/409.
The disc is constrained within $\sim$6 and 60 r$_s$.
The addition of a narrow absorption line improves the fit ($\chi^2$/d.o.f.=360/407)
with no appreciable change in the model parameters: although the best fitting value is different 
 the diskline parameters remain in good agreement with the values quoted before. 
 
When the line is fitted with a  {\small LAOR} line model, the  fit is equally good ($\chi^2$/d.o.f.=374/409) with very similar parameters: E=6.93$^{+0.77}_{-0.27}$~keV, EW=320$^{+170}_{-100}$~eV and the inclination 
{\it i} is constrained to be less than 33$^{\circ}$.
Adding the narrow absorption line yields the same improvement ($\chi^2$/d.o.f.=360/407). 
There is a strong indication that an ionised disc line is preferred 
with respect to a neutral  line. However, these fits are not conclusive as the 
presence of a disc line implies a reflection component in the spectrum.
This will be tested extensively in section \ref{subsec:reflection_335}.   


\subsection{Reflection continuum from the accretion disc}
\label{subsec:reflection_335} 

We  consider here  the role played by
reflection in the continuum emission formation. 
Four spectral models are tested in the following (see Table \ref{tab:ion_ref_335} for a summary). In all four models  we set  the outer radius of the disc to be {\it r$_{out}$}=400 r$_g$.
The inner radius is set to be  equal to the innermost stable orbit i.e. 1.24 ~r$_g$ in models  where a
Kerr metric is used (A, B, C) and 6r$_g$ in the model adopting the Schwarzschild metric (model D). Where possible, however, we have tried to constrain the inner radius by treating it as
a free parameter.
 All the models comprise an input primary  power law and a reflection component  for the continuum. Model A and B comprise also a Gaussian  emission line to reproduce the broad  Fe K$\alpha$, whereas in models C and D the line is included in the computation of the reflection spectrum. In models A, B and C the relativistic effects due to the strong gravity of the black hole
are taken into account by applying the relativistic blurring code {\small KDBLUR} (Fabian et al. 2002).
The code adopts the Kerr metric (Laor1991) and  it smears out the spectrum  according to the same  4 parameters described in the previous section for the disc lines: {\it q}, {\it r$_{in}$}, {\it r$_{out}$} and {\it i}. The {\small KDBLUR} code is applied to the reflection continuum and the emission line only.
To account for the absorption feature at $\sim$5.9 keV, an absorption line with the width fixed to 1~eV and  Gaussian profile is included in all the models;  physical models and a discussion of the nature and origin of this feature are deferred to the following section. 

\begin{table*}
\begin{minipage}{14cm}
\begin{center}
\caption{\label{tab:ion_ref_335}Best fitting parameters for the reflection models described in section \ref{subsec:reflection_335}. Relativistic effects and a narrow absorption Gaussian line at $\sim$ 5.9 keV are included in all of them. The parameters with no errors have been frozen.}
\begin{tabular}{c c c c c c c c c c }
  \hline\hline
\\
Model & $\Gamma$ & R & $\xi$ & E$_{broad}$ & EW$_{broad}$ & {\it i} & r$_{in}$ &  {\it q} & $\chi^2$/dof \\
      &          &   &  (ergs cm s$^{-1}$)  & (keV) & (eV) & (degrees) & r$_g$ & - &   \\   
\hline
\\
(A):  PEXRAV+ PLAW+  & 2.27$^{+0.08}_{-0.03}$ & 1 & - & 6.4 & 570$^{+350}_{-200}$  & $>$80 & 1.24 & 2.27$^{+0.53}_{-0.98}$ & 355/403  \\
2 GAU + KDBLUR \\
 \\  
(B): PEXRIV +PLAW+   &  2.30$\pm$0.05 & 1 & 1000  & 6.97 & 300$^{+390}_{-220}$ & $<$ 30 & $>$4 & $>$2.5 & 360/402   \\
2 GAU + KDBLUR \\
\\
(C): REFLION + &     2.20$\pm$0.02 & - & 760$^{+240}_{-160}$ & - & - & 25$\pm$10 & 1.24  & 1.8$^{+0.7}_{-0.5}$ & 344/402     \\ 
GAU + KDBLUR \\
\\
 (D): XION + PLAW+ & 2.17$\pm$0.03 & - & -& -& - & $<$27 & 6 & - & 355/403 \\
 GAU \\
\hline\hline
\end{tabular}
\end{center}
(A, B): R=$\frac{\Omega}{2\pi}$, where $\Omega$ is the solid angle of the reflecting slab subtended at the X-ray source \\
(D) The ``lamppost" geometry is assumed for {\small XION} (see text)
\end{minipage}
\end{table*}

We start with the {\small PEXRAV} model (Magdziarz \& Zdziarski 1995) which describes
the reflection of an incident power law by a slab of neutral gas  (model A in table~\ref{tab:ion_ref_335}). We initially fixed the emission line energy to be 6.4 keV  and the reflection fraction to be $R=1$ ($R$ represents the solid angle of the reflector $R=\Omega/2\pi$). In this fit the power law gets steeper,  with a photon index of 2.27$^{+0.08}_{-0.03}$, as expected when adding a reflection component to a simple power law model (Nandra \& Pounds, 1994). This model  provides a  good fit ($\chi^2$/d.o.f.=355/403) to the data, 
but it requires  viewing the disc at an unusually high inclination ($>$80$^{\circ}$). This
seems unlikely for two reasons. Firstly, such a high inclination is not expected for a
Seyfert 1 galaxy such as this. Secondly, and perhaps more importantly, at high inclinations
very weak reflection features are expected (George \& Fabian, 1991), whereas we find a very strong  Fe K$\alpha$  here ($\sim$ 500~eV).
  If we test this fit with an inclination angle fixed to 30$^{\circ}$, 
the $\chi^2$ worsens ($\chi^2$/d.o.f.=363/404), but it is still acceptable. However, we regard it more probable that the excess ``blue" flux is a consequence of the disc being ionised since
this interpretation is more consistent with the line energy (see Section~\ref{subsec:diskline}).

As a first attempt  we use the {\small PEXRIV} model (model B in table~\ref{tab:ion_ref_335}), which is basically 
a {\small PEXRAV} component with the disc temperature and the ionisation parameter of the gas as additional  parameters.
The ionisation parameter is the key discriminant when fitting ionised accretion disc spectra.
It is  defined as $\xi$=$\frac{4\pi F}{n_h}$ where F is the incident power law flux in 
ergs~s$^{-1}$~cm$^{-2}$  and n$_h$ is the hydrogen number density of the slab
in cm$^{-3}$. The spectral features in the resulting spectra depend on the ionisation state of the disc (e.g. Matt, Fabian \& Ross 1993, 1996).  
During the fitting procedure  the line energy is fixed at 6.97~keV for H-like iron, as indicated by the fits in section 
\ref{subsec:diskline}. 
 With the line being emitted at the maximum ionisation state, $\xi$ must be in the  range 500$<$$\xi$ $<$5000 ergs~cm~s$^{-1}$ and so it is fixed at 1000 ergs~cm~s$^{-1}$ (Matt, Fabian \& Ross 1993). 
 The emissivity index appears quite steep ({\it q}~$>$2.5) implying 
that the emission is concentrated in the inner regions. 
The inclination in this fit seems more plausible, being constrained to be $<$30$^{\circ}$. 


We next employ  the ionised disc model described in Ross \& Fabian (2005)  ({\small REFLION}, model C in table~\ref{tab:ion_ref_335}), based on the models 
first proposed  by Ross \& Fabian (1993)  and Ross et al. (1999).
This model calculates the reflection spectrum from an infinite slab of ionised gas illuminated by an X-ray power-law continuum. All the ionisation states and transitions are included in the calculated spectrum.
The ionisation state of the slab, the spectral index and the iron abundance are free parameters.
The best fitting parameters for this model  are listed in Table \ref{tab:ion_ref_335}, model C.
The fitted  iron abundance is not listed in the table but it is close to the Solar value (Fe/solar$\sim$0.9).


\begin{figure}
\begin{center}
\psfig{figure=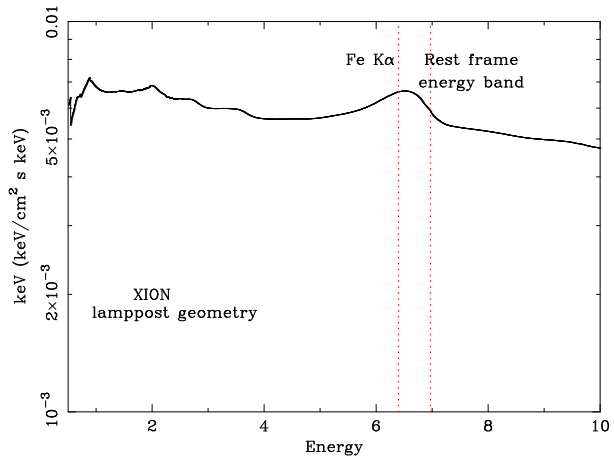,height=6cm,width=8cm}
\caption{\label{fig:xion_335}Rest frame plot of  the model XION (lamppost geometry). The full energy range of the rest frame Fe K$\alpha$ line emission is highlighted between the dotted lines: the relativistic smearing is very clear, since the line emission is blurred over 
the continuum, resulting in a broad and smooth hump.} 
\end{center} 
\end{figure}

The computation of the reflected spectrum assuming constant density in the reflecting slab is not entirely  physical;
as studied in Nayakshin, Kazanas \& Kallmann (2000),  thermal and ionisation  instability in the disc will lead to the formation of a top layer of ionised material, superposed  on cold neutral gas. 
In such a case, the density varies along with the ionisation state and radiation field, and  it cannot be 
considered constant.
Therefore, we decide to fit our data with the  {\small XION}  model  (model D in table~\ref{tab:ion_ref_335}) developed by Nayakshin \& Kallman (2001) and implemented in {\small XSPEC}, which calculates  the reflected spectrum in hydrostatic balance, taking into account the ionisation instability in the disc.  
After choosing one of the available geometries, the model parameters are the distance between the disc and the source of X-ray photons, the accretion rate, the luminosity 
of the X-ray source, the inner and outer disc radii and the spectral index.
The reflected spectrum is calculated for $\sim$30 different radii 
and then integrated over the disc surface. Relativistic smearing 
is included for a non-spinning black hole  according to Fabian et al. (1989). 
{\small XION} includes three types of geometry:  a compact X-ray source located 
above the disc (the so-called  lamppost), a  central sphere with an 
outer cold disc and  a  magnetic flares geometry.
All of them have been tested without finding any discriminant, so we resolved to assume
the simplest configuration (lamppost). The model is shown in Fig.\ref{fig:xion_335}. 
 Assuming  that the reflecting  disc extends between 6 and 400~r$_g$, we get the following  parameters:
 $\Gamma$=2.17$^{+0.03}_{-0.03}$
 and  the inclination of the disc is consistent with a plausible value of $\sim$ 30$^{\circ}$ (see Table \ref{tab:ion_ref_335}).
 The height of the X-ray source above the disc is constrained to be lower than 10~r$_g$.
The model is statistically comparable to the others  in  Table \ref{tab:ion_ref_335},
yielding  $\chi^2$/d.o.f.=355/403.

Clearly, ionised disc models are able to fit the spectrum of Mrk 335 rather well.  
 Nonetheless,  we test an alternative scenario in the next section. 
\subsection{Alternative to disc reflection:  partial covering}
\label{subsec:part_cov_335}
\begin{figure}
\begin{center}
\psfig{figure=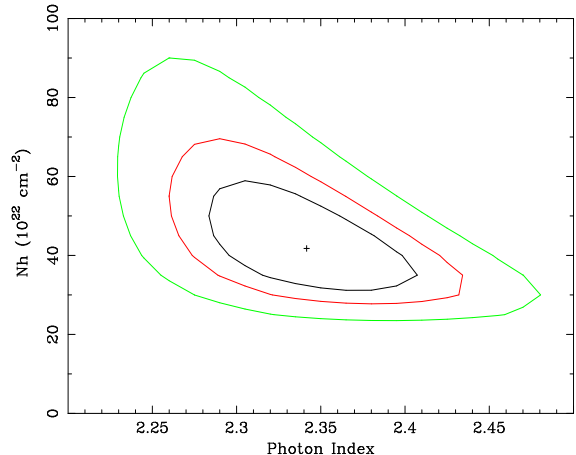,height=6cm,width=7.5cm}
\caption{\label{fig:nh_335}Confidence regions for the power law slope and the column density of the cold absorber in the partial covering model described in section  \ref{subsec:part_cov_335}.}   
\end{center} 
\end{figure} 
The spectral curvature in Mrk~335  could be the result of cold gas partially covering the 
source of primary photons.
This hypothesis is tested by fitting the spectrum with a model 
expressed by A(E)*M(E) where A(E) is the primary power law and the  absorption component  M(E)  is defined as 
\begin{displaymath}
M(E)=C_f e^{-N_H \sigma(E)} +(1-C_f)
\end{displaymath}  
The absorption component depends on the column density of the absorbing material N$_H$,
on the  photoelectric absorption cross section $\sigma$ and on the fraction of the source 
obscured by the absorber,  C$_f$.
Such a model is implemented in {\small XSPEC (PCFABS)} and it yields an acceptable fit to the data,  with $\chi^2$/d.o.f.=373/407.
The best fit parameters are consistent with a steep power law, $\Gamma$=2.34$^{+0.06} _{-0.06}$.
The covering fraction  is found to be   38$^{+7} _{-7}$ percent.
The column density of the medium is found to be   4.1$^{+2.0} _{-1.1}$ 
$\times$10$^{23}$ cm$^{-2}$.
The confidence regions are shown in Fig.\ref{fig:nh_335}, and indicate that over a large range of photon indeces,  the column density is  of this order of
magnitude.
Adding a narrow  Gaussian line with negative intensity  to the {\small PCFABS} component   improves the fit by $\Delta\chi^2$=10 for 2 free parameters.
The energy is again very tightly constrained to 5.92$^{+0.03} _{-0.05}$~keV.
The $\chi^2$/d.o.f. (363/405) is comparable with those found in the best fitting reflection models.
The presence of a column of cold gas shielding a photoionizing  source,  
implies  that a fluorescence  line is emitted  at 6.4 keV,  following the absorption. 
We have checked the 90\% upper limit of the EW for  a neutral line with $\sigma$=~0.01 keV and it is  found to be $<$45 eV. 
For a spherical absorber fully covering the source, the expected EW for a column density of $3\times$ 10$^{23}$~cm$^{-2}$ is approximately 100 eV (Leahy \& Creighton 1993). With the best-fit covering fraction above, we therefore expect the EW to be $\sim 40$~eV, just consistent with the upper limit quoted above. The partial covering model cannot therefore be ruled out on the basis of the lack
of a line core.


\section{The narrow absorption line}
Regardless of the model of the broad emission around iron-K in Mrk 335, we find evidence 
for a narrow notch-like feature at 5.9 keV detected in the EPIC data. The significance of this feature implied by a standard F-test, is higher than 99.73 per cent. As it has recently been shown by Protassov et al. (2002), the F-test may not be a reliable indicator of the significance of narrow feature such as these. It is therefore necessary to test the significance using simulations.

\subsection{On the significance of the narrow absorption feature}
\label{subsec:signif_335}
 To test the significance of the absorption feature we have performed Monte Carlo simulations assuming two baseline models (Table \ref{tab:baseline_335}).
The first  consists of a power law + broad Gaussian line (as described in section \ref{sec:spec_an_335}) while the second  is a power law + {\small DISKLINE}  (as described in section \ref{subsec:diskline}).   
The values of $\Delta\chi^{2}$ for adding the narrow line are 16 and 13 respectively for the gaussian 
lines baseline and for the diskline baseline.
We  test the null hypothesis, namely  ``what is  the probability that  a value of $\Delta\chi^{2}$  this large or larger will be obtained by chance?".
To answer this question, 10000 spectra  have been simulated with {\small XSPEC}   assuming the baseline model {\it without} the absorption line,  folding it  through the same instrumental response and adding noise randomly.
These spectra  have then been grouped according to the same criterion adopted 
for the real data set, i.e. 50 counts per spectral bin. 
In this way,  10000 fake background-subtracted data sets have been obtained with photons statistics corresponding to 30~ks  in the pn detector (around 28000 counts between 2 and 10 keV).
Each of these spectra is fitted with the baseline model.
 Then, a narrow absorption line is added to the fit, and
the improvement in  $\chi^2$ with respect to the baseline model is determined. 
The width of the absorption line is fixed to 1~eV during the fitting and 
 the energy of the line is stepped  in increments of 70~eV.
 To avoid any effect of the calibrations at  the boundary of the instrumental response, the search is performed in the energy range 2.5--9.5 keV.  
  This process was performed for both baseline models.
 We obtain a significance of 99.70 per cent  for the gaussian baseline model which confirms the goodness of the detection. The value obtained using the {\small DISKLINE} model 
 is slightly lower, but still strongly suggests that the null hypothesis can be rejected.



\begin{table}
\begin{tabular}{c c c }
 \scriptsize 
  \\ \hline\hline
 Baseline  & F-test sig. & M.Carlo sig. \\
 Model    &            -        &          -               \\
 \hline
  P. law+ broad Gau     &   $>$99.73\% &       99.70\%  \\
  E=6.49 keV, $\sigma$=0.4 keV &            \\
  \hline 
  P.law+Diskline      & $>$99.73\%  &      98.44\%           \\
  E=6.99 keV, {\it i}=21$^{\circ}$, q=3  &    \\ 
  \hline\hline
 \end{tabular}
 \caption{\label{tab:baseline_335}Baseline models used as input models in the Monte Carlo simulations to test the robustness of the detection as found by the F-test.}
\end{table}


\subsection{The narrow feature: complex absorption}
\label{subsec:narrow_complex_335}
\begin{figure*}
\begin{center}
\centerline{
\resizebox{8.5cm}{!}{\psfig{figure=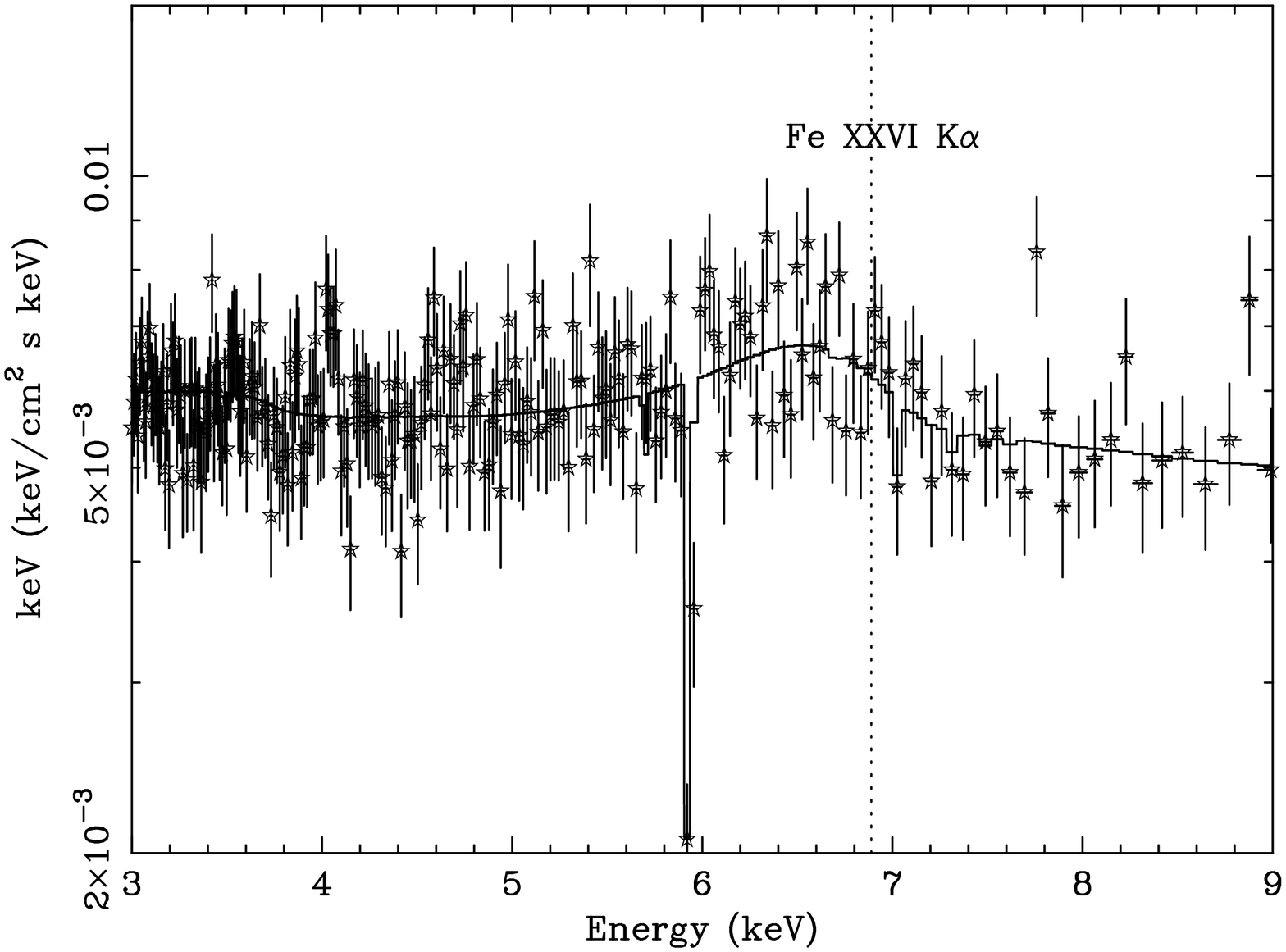}}
\hspace{0.2 cm}
\resizebox{8.5cm}{!}{\psfig{figure=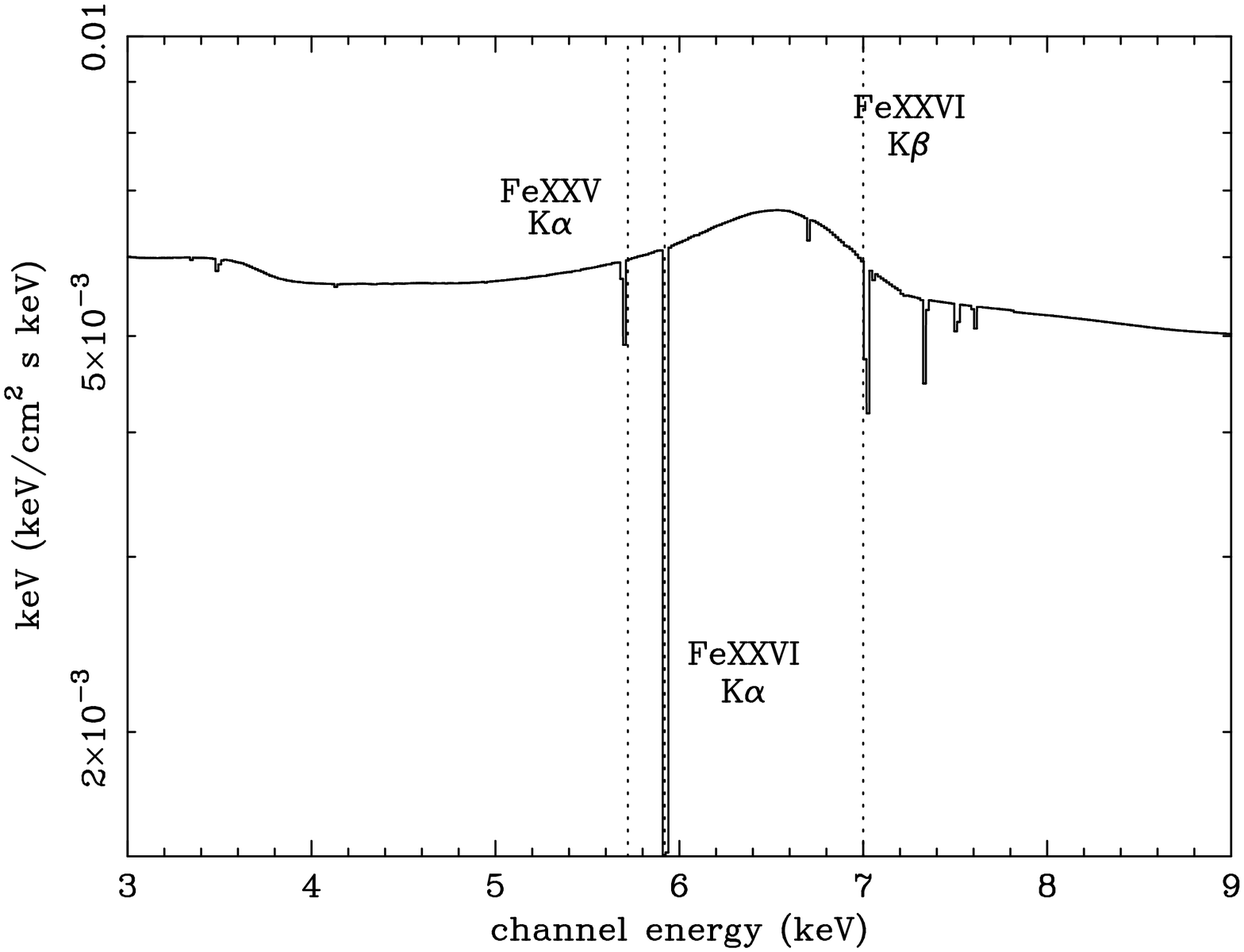}}
}
\caption{\label{fig:xstar_models_335}{\it Left panel}:
Rest frame plot of the 3-9 keV data fitted with the best fitting  complex absorption model {\small XION+XSTAR} described in \ref{subsec:narrow_complex_335}. The energy of the resonant Fe XXVI K$\alpha$ line is labelled to emphasize the energy shift of this line, clearly detected at $\sim$ 5.9 keV. {\it Right panel}: Rest frame plot of the model used to fit the data on the left. The combination of the effects of the reflection spectrum and the highly photoionised   gas is visible: the first one reproduces the spectral curvature  between 5 and 7 keV, the latter imprints redshifted absorption features on the continuum.  
In our data, we observe the Fe XXVI K$\alpha$ only, the other ones being too weak.}   
\end{center} 
\end{figure*}

The most obvious identification for the absorption feature is with redshifted iron K$\alpha$ resonance absorption. The identification with iron is favoured since the observed energy of the line is too high to be readily explained by K$\alpha$ absorption in any of the other astrophysically abundant elements. Furthermore, the absence of any related absorption lines at softer energies (Gondoin et al. 2002) suggests a high ionization state for the absorbing material. Thus we consider there to be two plausible identifications, with iron xxv K$\alpha$ or xxvi K$\alpha$.
  It may be possible to test between these by modeling the absorption line using a photoionisation code. We
therefore used {\small XSTAR} to generate a grid of photoionisation models with
various ionisation parameters and column density.
Solar elemental abundances and turbulence velocity of 100 km/s were assumed.
This grid has been incorporated into {\small XSPEC} as a table model
with 3 additional free parameters: i) the column density (named N$_w$ to be distinguished 
from the cold column in Table~1);  ii) the ionisation 
parameter $\xi$;
iii) a redshift parameter  which includes  all possible contributions (i.e. the cosmological redshift ({\it z$_{source}$}), the velocity of the absorber  ({\it z$_{inflow}$}) and the  gravitational redshift  ({\it z$_{grav}$})).

 We have tested a model in which  the Gaussian absorption line in model D of Table~\ref{tab:ion_ref_335} is replaced by this {\small XSTAR} model. The results are plotted in the left panel of  Fig. \ref{fig:xstar_models_335} and the fit  yields $\chi^2$/d.o.f.=359/401.  
Four main absorption features  are imprinted on the continuum (see right panel of Fig. \ref{fig:xstar_models_335}): 
the K$\alpha$ and K$\beta$  transitions of Fe {\sc xxv} and  {\sc xxvi}. However, only
the  K$\alpha$  ones are sufficiently strong to be important here.
The energies of the transitions of interest are reported in Table \ref{tab:resonant_lines_335}. 
The absorption line detected in the data is consistent with identification with the Fe {\sc xxvi} K$\alpha$ line, whereas the other features predicted by the model are too weak to be detected at the CCD resolution. If this identification is correct, the redshift correponds to an inflow  velocity of $\sim$~0.15~{\it c}, if gravitational redshit is neglected.

\begin{table}
{\small  ABSORPTION LINES}\\
\begin{tabular}{c c c c c}
\\ \hline \hline
Ion & Transition & E$_{\it em}$ & E$_{\it obs}$ &  inflow {\it v}  \\
\hline
Fe XXV &  K$\alpha$      & 6.70 keV   & -   & -  \\
       &  K$\beta$  & 7.88 keV & - & -  \\  
\\
Fe XXVI &  K$\alpha$  &  6.97 keV &  5.92 keV &  0.15 {\it c} \\  
        &   K$\beta$     & 8.17 keV & - &-    \\
\hline \hline
\end{tabular}
\caption{\label{tab:resonant_lines_335}The ionisation state in our XSTAR model allows to have only He-like and H-like Fe ions; in this case   the  Fe K complex consists of the ions described in the table. The rest frame energies are specified for all lines. We include  the observed energy (corrected for the source redshift)   and the  corresponding inflow velocity  for the only detected line.}
\end{table}


\section{A simple model for inflow}
\label{sec:model_sim_335}

To elucidate the inflow hypothesis, in this section we present a simple physical model
and use it to synthesise X-ray spectra for qualitative comparison with the
observed absorption feature. 

\subsection{Flow structure}
\label{subsec:flow_structure_335}

Given the modest data quality and that only 
one line is detected, we will consider only the simplest class of inflow 
models,
that of spherically
symmetric radially infalling gas. 
We assume that the gas occupies a region which extends from inner radius $r_{\mbox{\scriptsize in}}$ to outer radius $r_{\mbox{\scriptsize out}}$
from the central black-hole. 
We adopt a velocity-law for the gas which is appropriate for radial infall allowing for the possibility of significant 
repulsive radiation pressure, namely

\begin{equation}\label{eq:velocity_335}
v(r) = - c \sqrt{ 2r_{g} / r} \sqrt{ 1 - L / L_{\mbox{\scriptsize edd}}}
\end{equation}
where $r_{g} =  G M_{\mbox{\scriptsize BH}} / c^2$ is the gravitational radius of the black-hole, $r$ is
the radial co-ordinate and $L / L_{\mbox{\scriptsize edd}}$ is the ratio of the luminosity of the central region to the
Eddington luminosity for the central black-hole. 
The last term in equation~\ref{eq:velocity_335} accounts for the effect of repulsive radiation pressure on electrons only - it neglects radiation pressure
due to spectral lines. This is likely to be a good assumption since it is found in all the models discussed below that the gas is very
highly ionised.
For the black-hole in Mrk~335, we adopt a mass 
$M_{\mbox{\scriptsize BH}} = 1.4 \times 10^7$~M$_{\odot}$ (Peterson et al., 2004).

To specify the density of the flow we use a prescription based on a constant mass infall rate ($\Phi$), namely

\begin{equation}\label{eq:density_335}
\rho (r) = \frac{\Phi}{4 \pi r^2 v(r)}
\end{equation}
In addition, we have fixed the electron temperature of the flow at $T_{e} = 10^6$~K, close to the primary 
black-body temperature temperature fit to the observed spectrum (Gondoin et al. 2002). Solar abundances are adopted.
\subsection{Spectral synthesis method}

With the flow model described above, spectra were synthesised for
the 2 -- 10~keV region using a Monte Carlo radiative transfer code based closely on that
discussed by Sim (2005) which uses methods developed by Lucy (1999, 2002, 2003).
The code includes the effects of spectral lines, bound-free edges and Compton scattering
by free electrons (although inverse Compton scattering is not included).
The atomic data used for the calculations, and the 
treatment of excitation and ionisation of the flow are exactly as discussed by
Sim (2005).

The radiation field incident on the lower boundary of the flow consists of the same 
three components as used by Sim (2005): a multi-colour black-body disc (following 
Mitsuda et al. 1984; the particular implementation is exactly as used by
Sim 2005), a soft-excess (modelled as a black-body) and
a power-law tail extending to X-ray energies.
For the computation presented here, the 
power-law photon index was fixed at $\Gamma = 2.2$ (based on fits to the data,
see Section \ref{sec:spec_an_335}) and the 
normalisation fixed by
requiring the total 2 -- 10~keV X-ray flux to match the observed luminosity of
$1.8 \times 10^{43}$~ergs~s$^{-1}$.
The disc luminosity was set to 60 per cent of the Eddington luminosity for the central black-hole, 
close to the value $L / L_{\mbox{\scriptsize edd}} = 0.62$ reported for Mrk~335  
by Gierli\'{n}ski \& Done (2004) based on the work by Boroson (2002).
The soft-excess was modelled as a black-body with temperature $1.3 \times 10^6$~K 
and normalisation fixed to the power-law component.

In contrast to that presented by Sim (2005), the code used here includes the full
special relativistic expression for the Doppler shift and approximately accounts for
gravitational redshift using

\begin{equation}\label{grav_red_335}
\gamma (1 - \mu v(r) / c) \nu = {\nu^{\prime}} \sqrt{ 1 - 2r_{g}/r}
\end{equation}
where $\nu^{\prime}$ is the frequency of a photon at radius $r$ as measured in the comoving
frame, $\mu$ is the usual direction cosine, $\gamma = ({1 - v^2/c^2})^{-1/2}$ and
$\nu$ is the photon frequency that would be recorded by an infinitely distant observer 
at rest relative to the black-hole. Other relativistic effects (including aberration of
angles) are neglected.

Monte Carlo simulations were performed to determine the ionisation state and 
provide the necessary 
estimators for the radiation field before synthesising a spectrum by tracing rays
as described by Lucy (1999) and Sim (2005). The method was generalised to account for
cases where photon frequencies blue-shift in the co-moving frame along parts of their
trajectories (Lucy 1999 and Sim 2005 consider only accelerating outflows in which
the co-moving frequency decreases monotonically along all photon paths).
\subsection{Results from the model}

\subsubsection{Inflow extending over a wide range of radii}
\label{subsubsec:inflow_largeradii_335}

We begin by showing that the data can rule out
the simplest inflow model that can be considered, that of a smooth continuous flow which extends from near ($\sim$ several $r_{g}$) 
the central black-hole to much greater distances. Fig.\ref{fig:broad_pcyg_335} shows the 3 -- 9 keV spectrum computed for a model of this
sort with $r_{in} = 20$~$r_{g}$, $r_{out} = 2\times~10^3$~$r_{g}$, $\Phi = 0.2$~$M_{\odot}$~yr$^{-1}$ and neglecting radiation
pressure in the assumed velocity law (i.e. setting $L / L_{\mbox{\scriptsize edd}} = 0$ in equation \ref{eq:velocity_335}). In agreement with the observations, this model
predicts very few spectral features since the ionisation state of the gas is very high. The dominant feature is a broad
inverse P~Cygni Fe~{\sc xxvi} K$\alpha$ line, but a weaker feature due to the K$\beta$ line of the same ion appears at harder
energies.

\begin{figure}
\begin{center}
\psfig{figure=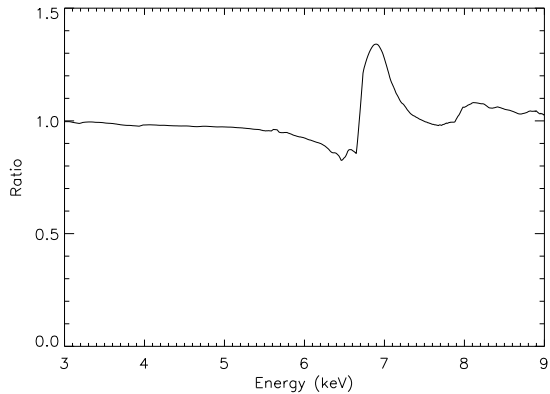,width=8cm}
\caption{\label{fig:broad_pcyg_335}Computed 3 -- 9 keV X-ray spectrum for an inflow with $r_{in} = 20$~$r_{g}$, $r_{out} = 2\times~10^3$~$r_{g}$, $\Phi = 0.2$~$M_{\odot}$~yr$^{-1}$, neglecting radiation pressure. The plot shows the ratio of the computed flux to a power-law with index
$\Gamma = 2.2$. The energy is given in the quasar rest frame. The absorption features are predominantly due to Fe~{\sc xxvi} K$\alpha$ (around 6.5 keV) and
K$\beta$ (around 7.7 keV).}
\end{center} 
\end{figure}

However, although the model predicts the correct absorption line, it
can be strongly ruled out by 
the data since the line shape and position are quite different: the observed feature lies at $\sim$5.9~keV and is sufficiently 
narrow to be unresolved in the data. 
In contrast, the computed absorption line is very broad, extending from
around 5.5~keV to a point of deepest absorption at $\sim 6.5$~keV. 

The large line width in the model is a result of the large radial extent, and therefore large velocity range, adopted for 
the flow. Fig.~\ref{fig:ion_frac_335} shows the Fe ionisation state of the flow as a function or radius. At all points, the fully ionised
state (Fe~{\sc xxvii}) is dominant with a significant contribution from the H-like ion (Fe~{\sc xxvi}). The He-like
stage does not account for more than a few percent anywhere in the model and thus it is the H-like ion that dominates the line 
formation. There is a gradient of ionisation present such that a higher ionisation state (i.e. less Fe~{\sc xxvi}) occurs at small
radii. For the density stratification given by equation~\ref{eq:density_335}, this ensures that the Fe~{\sc xxvi} line
opacity remains significant out to large radii which gives rise to the low-velocity absorption seen in the spectrum.
For the velocity and density laws specified by equations \ref{eq:velocity_335} and \ref{eq:density_335}, to obtain a line profile closer to that observed 
(i.e. in which the opacity is concentrated at high velocities) would require an ionisation gradient in the opposite sense to that
obtained here -- however, that cannot be readily achieved since the influence of self-shielding in the flow will always 
favour a decreasing ionisation state with radius. Thus, in order to obtain a model which predicts spectra closer to that observed,
one must consider alternative models to the continuous, large-scale spherical inflow.

\begin{figure}
\begin{center}
\psfig{figure=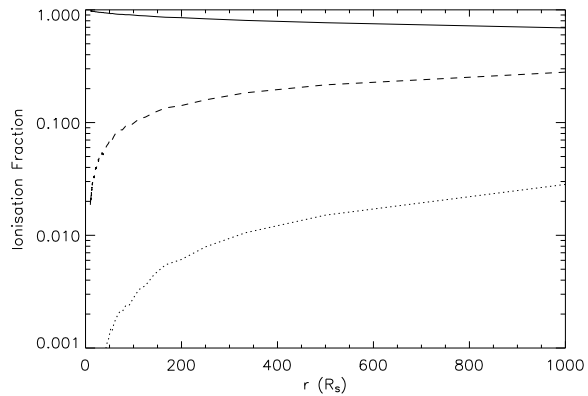,width=8cm}
\caption{\label{fig:ion_frac_335}Computed Fe ionisation fractions as a function of radius for the same flow discussed
in Section \ref{subsubsec:inflow_largeradii_335}. The curves represent Fe~{\sc xxvii} (solid line), Fe~{\sc xxvi} (dashed line) and
Fe~{\sc xxv} (dotted line).}
\end{center} 
\end{figure}

\subsubsection{Inflow extending over a limited range of radii}
\label{subsubsec:inflow_smallradii_335}
As discussed above, a large-scale spherical inflow model predicts an Fe~{\sc xxvi} absorption line which is too broad for consistency 
with the data. However, consistency can be obtained by restricting the range of radii in which the density of infalling gas is
significant. Such a model might approximately describe a discrete infalling
blob of gas or a larger scale inflow in which the accretion rate is variable leading to sections of the flow which are significantly more dense
that others.
With the model considered here, this can be readily achieved by limiting the radial range occupied by the flow.
In order that the deepest absorption occurs at around $5.9$~keV, corresponding to a redshift of the Fe~{\sc xxvi} line
by 0.15c, 
a strong constraint can be placed on the outer boundary of the inflow:
$v(r_{\mbox{\scriptsize out}}) \approx 0.15c$ is required. Using equation \ref{eq:velocity_335}, this leads to $r_{\mbox{\scriptsize out}} = 100$~$r_{g}$ if
the radiation pressure is neglected. When radiation pressure is accounted for with $L / L_{\mbox{\scriptsize edd}} = 0.6$, the 
required outer radius is smaller, $r_{\mbox{\scriptsize out}} \approx 48$~$r_{g}$. Fig.~\ref{fig:narrow_inflow_335} shows the computed spectrum
for a flow with $r_{\mbox{\scriptsize out}} = 48$~$r_{g}$, $r_{\mbox{\scriptsize in}} = 24$~$r_{g}$ and
$\Phi = 0.3$~$M_{\odot}$~yr$^{-1}$. This model provides a good description of the narrow absorption line identified in the data.
Note that the model does not include the formation of the Fe~K$\alpha$ emission line which was discussed in Section~\ref{sec:spec_an_335}. This most probably accounts for the excess emission to the blue of the absorption feature.

The constraint on the inner radius adopted for the model is less strong than that on the outer radius 
since the opacity drops at high velocities due to the combination of higher velocity gradient, higher ionisation state and shorter 
path length. Thus the adopted value of 
$r_{\mbox{\scriptsize in}} \approx 24$~$r_{g}$ should only be regarded as an upper limit on $r_{\mbox{\scriptsize in}}$. 

\begin{figure}
\begin{center}
\psfig{figure=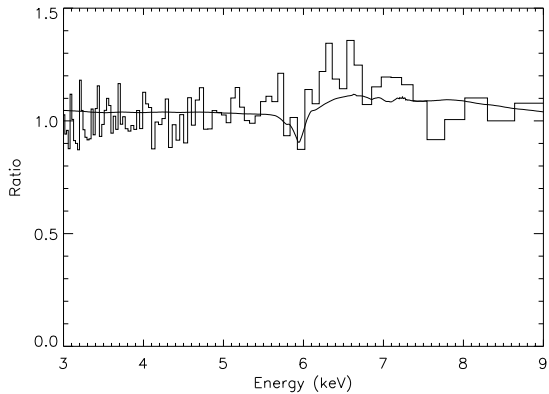,width=8cm}
\caption{\label{fig:narrow_inflow_335}Computed 3 -- 9 keV X-ray spectrum for an inflow with $r_{in} = 24$~$r_{g}$, $r_{out} = 48$~$r_{g}$, $\Phi = 0.3$~$M_{\odot}$~yr$^{-1}$,  accounting for radiation pressure due to electrons at 60 per cent of the Eddington limit. The plot shows the ratio of the computed flux to a power-law with index
$\Gamma = 2.2$. The energy is given in the quasar rest frame. The absorption feature is due to Fe~{\sc xxvi} K$\alpha$. The light histogram shows the observational data for Mrk~335, also normalised to a power-law fit.}
\end{center} 
\end{figure}

The mass accretion rate in the flow, $\Phi = 0.3$~$M_{\odot}$~yr$^{-1}$, was 
chosen in order to obtain an absorption line strength consistent with the
observations. This value can be compared with the Eddington accretion rate
for the central black-hole, given by

\begin{equation}
\Phi_{\mbox{\scriptsize Edd}} = \frac{4 \pi G M_{\mbox{\scriptsize BH}}}{\eta \kappa c}
\end{equation}
where the $\kappa$ is the opacity and $\eta$ is the efficiency of the accretion process in producing radiation. For our adopted 
$M_{\mbox{\scriptsize BH}} = 1.4 \times 10^7$~M$_{\odot}$ with
$\kappa = 0.86 \sigma_{T} / m_{H}$ (where $\sigma_{T}$ is the Thomson cross-section)
and adopting $\eta = 0.07$ this gives
$\Phi_{\mbox{\scriptsize Edd}} \approx 0.5$~$M_{\odot}$~yr$^{-1}$.
Thus our model inflow has $\Phi / \Phi_{\mbox{\scriptsize Edd}} \approx 0.6$, 
in 
remarkable
agreement with our observationally motivated assumption that 
$L / L_{\mbox{\scriptsize Edd}} = 0.6$. The value of $\Phi$ needed by the model is moderately well constrained. For lower values of $\Phi$ (below about 
0.15~$M_{\odot}$~yr$^{-1}$) the predicted absorption line is too weak to be 
detectable; this is a strong effect because reducing $\Phi$ 
both decreases the Fe density and makes the gas more highly ionised.
It is unphysical to consider steady-state spherical models with 
significantly higher values of $\Phi$ since 
$\Phi / \Phi_{\mbox{\scriptsize Edd}} > 1$ is unacceptable for such a model.
However, this limit can be avoided if the 
inflow subtends only a fraction $b$ of 
solid angle since in such a case the true
accretion rate onto the black-hole would be only $b \Phi$.
Although we have not yet extended our modelling to consider non-spherical 
geometries, the absorption part of the line profile computed from the spherical
model is likely to remain approximately valid for a non-spherical inflow which
lies directly in the line of sight of an observer.
Nevertheless, the data excludes values of $\Phi$ significantly greater than 
that adopted in the model since for higher values of $\Phi$, the line becomes stronger and 
Compton downscattering in the flow causes the continuum to bend downwards
at high energies.


\section{Discussion}
\label{subsec:discu_broad_335}
We have performed a detailed investigation of the hard X-ray spectrum of Mrk~335 as seen by {\it XMM-Newton}. The main results can be summarised as follows.
\begin{itemize}
\item The presence of a broad  Fe K line associated to 
a reflection component is confirmed, as previously reported in Gondoin et al. (2002).
This line is well fitted by disc  reflection models, as long as the accretion disc is
strongly photoionised. Nonetheless, a partial covering component provides a statistically
plausible alternative for  the broad residuals.
\item For the first time, a narrow absorption line is detected at $\sim$5.9~keV.
The significance of this line based on Monte Carlo simulations is 99.7~per cent. If real, this feature
may be the signature of infalling gas in Mrk 335. 
\item A model for the inflow has been developed and qualitatively compare to the data. We
find that the observed feature can be matched if it is identified as Fe {\sc xxvi} in highly ionised gas,
inflowing as at a few tens of gravitational radii. 
\end{itemize}

\subsection{On the interpretation of the broad residuals}

Mrk 335 presents clear evidence for broad emission from $\sim 5-7$ keV. Such emission has been found to be common in the X-ray spectra of AGN observed by ASCA  and interpreted as relativistic emission from an accretion disc (Nandra et al. 1997). As we have shown, there is excellent agreement between such models and the data in the case of Mrk 335. This is particularly important because Mrk 335 is one of only a few bright Seyferts which show no evidence for photoionised gas in the line of sight (Blustin et. al 2004). This fact rules out alternative interpretations for the broad residuals based on complex absorption in high ionisation gas (Reeves et al. 2004). Furthermore, the lack of any 6.4 keV core to the line,
otherwise common in AGN (Yaqoob \& Padmanhaban 2004), rules out any significant contribution from a distant reflector such as the molecular torus (Ghisellini, Haardt \& Matt 1993; Krolik, Madau \& Zycki 1993). This removes another potential complication in the interpretation of the broad residuals. 

The only alternative to the disc line we have not been able to rule out with the present dataset is
that of partial obscuration by a neutral medium. While such a model is statistically as good as accretion disc reflection, and it has been applied in an {\it ad-hoc} way to model the spectra of active galaxies, there is little or no physical motivation for such a component. As the source(s) of X-ray emission in AGN are thought to be very small and close to the central black hole, to partially cover the source the absorbing material would also have to be very small and very close to the X-ray source. It could only remain neutral in these circumstances if it were of extremely high density. Such a scenario has been suggested by Guilbert \& Rees (1988).  In their model, however, the clouds are optically thick and would then be a source of relativistic line emission and Compton reflection similar to the accretion disc, rather than neutral absorption. This cold cloud model has been explored further by Celotti, Fabian \& Rees (1992) and Kuncic, Celotti \& Rees (1997) who suggested that there may be ultra-dense small clouds of lower column density in the central regions of AGN, which may be magnetically confined (Rees 1987). These would then represent a plausible site for the patchy absorption. Future observations of Mrk 335 at high signal-to-noise ratio should be able to rule out partial covering as an alternative to accretion disc reflection, either on pure statistical grounds, or by tightening the constraints on the expected line emission from such gas. For the purposes of the present discussion, however, we concentrate on the interpretation that the broad residuals arise from a relativistic accretion disc. 


\subsubsection{Reflection from an ionised  accretion disc}
\label{subsubsec:discu_ref_335}

The 2-10 keV spectrum has been fitted with a single Gaussian line and with  diskline models (see section \ref{subsec:diskline}).  The large width of the Gaussian line corresponds to a velocity of the emitting material  of 8.8 $\times$10$^{4}$ km s$^{-1}$. Such fast motion implies that 
the line emitting material is located at a small distance from the source of X-ray radiation
and therefore that the profile is modified by  relativistic effects due to the gravitational field. 

Significant emission is found both above and below the rest energy of  known Fe K$\alpha$ emission lines. The observation of redshifted emission is the classic signature of accretion disc lines (e.g. Tanaka et al. 1995) and furthermore it would be against any model in which the broad emission complex consists solely of a blend of narrow lines (e.g. Bianchi et al. 2003). The fact that we also see broad, blueshifted emission is of particular interest, as it tends to imply that the disc is either highly inclined to the line of sight, or highly ionised. The latter seems considerably more likely in the case of Mrk 335. We have demonstrated that neutral reflection models require an extremely high inclination ($>$ 80$^{\circ}$), unlikely for a type I Seyfert such as this and inconsistent, at least in the simplest interpretations, with the large EW of the line.

Several ionised reflection models have been tried and they provide a physical interpretation of the data.  They all converge to  indicate that the disc is fairly uniformly illuminated over a large range of radii, i.e. 1.24-6~$r_g$ to 400~$r_g$, and  that it is viewed at an angle of $\sim$ 30$^{\circ}$.
With the present data we cannot discriminate clearly between a spinning or non-spinning 
black hole. The inner radii in Table \ref{tab:ion_ref_335}  indicate a very small value 
for the inner edge of the reflecting area of the disc, but there is certainly no strong requirement for
emission within the last stable orbit of a Schwarzschild black hole. The Fe K$\alpha$  line peak energy is consistent with 6.97~keV, implying that it is emitted by  Fe  at the highest 
ionisation stage (Fe {\sc xxvi}). The line EW   measured in section \ref{subsec:diskline}   is found to be $\sim$  400~eV. Taking into account  the power law slope in Mrk~335,  this value is  in good agreement with the scenario predicted  by  Nayakshin  et al. (2000), who state that steep
X-ray spectra ($\Gamma$$>$2) may lead to the production of an intense and highly ionised line.



\subsection{Narrow absorption line} 
\label{subsec:discu_narrow_335}
The narrow absorption line at $\sim$~5.9~keV is certainly the most intriguing feature in  the present data. It is detected at $>$99~per cent confidence, and identified as a resonance feature of very highly ionised iron (Fe {\sc xxvi}), consistent with the lack of soft X-ray absorption features in the RGS
spectrum (Gondoin  et al. 2002) and of any other absorption in the EPIC data. 
The physical interpretation of the line
does not depend on the continuum interpretation, so 
it is discussed separately in this section.

If the effect of the  gravitational field is neglected and the energy shift is 
attributed to the velocity of the material, then  
the observed redshift of the line corresponds to a receding  velocity 
of 50,000~km~s$^{-1}$  in the absorbing gas. 
Therefore, this number represents an upper limit to the flow speed.
 If the distance of the absorber was known with sufficient precision, it would be possible to
disentangle  the contribution of the  gravitational shift more precisely.
Unfortunately, no information on the allowed range of radii can be extracted 
from the line width, being it unresolved by the EPIC CCDs.

However, the narrowness of the line indicates that the interpretation of the redshift as due to inflowing gas may be preferable to the scenario proposed by Ruszkowski \& Fabian (2000). In their model, an absorption Fe K line is produced in a rotating, rather than inflowing plasma in the vicinity of the black hole. The observed redshift of the line is then primarily due to gravitation, but the line profile is predicted to be significantly broadened, and indeed has a characteristic profile, as do accretion disc
emission lines. Higher spectral resolution is required to rule out this hypothesis definitively, but the present data show no evidence for such broadening. In fairness we should point out that the inflow model presented in section \ref{sec:model_sim_335}  also predicts a very broad line for a continuous flow, and is only
consistent with the data if we restrict the range of radii to be compatible with the narrowness of the
observed line.

In the recent literature,  there have been detections of redshifted and blueshifted absorption lines 
indicating the presence of  high velocity gas in the centres of Active Galaxies.
The detections of blueshifted absorption features
in quasar spectra have been interpreted as relativistic outflows of matter originating closely to the central 
black hole (PG1211+143 and PG0844+349 Pounds et al. 2003a, 2003b, PDS 456 Reeves et al. 2003,  APM 08279+5255 and  PG1115+080 Chartas et al. 2002, 2003).
We note, however, that  some of these claims may be explained  with an alternative hypothesis  of features due to local gas in the Galaxy rather than intrinsic to the quasars (McKernan et al. 2005). 
 
Most of the redshifted lines instead,  are unambiguously interpreted as resonant lines of highly ionised Fe originated in absorbers intrinsic to the  nucleus   (NGC 3516, Nandra et al. 1999, Mrk 509, Dadina et al. 2005, E 1821+643, Yaqoob \& Serlemitsos 2005, Q0056-363, Matt et al. 2005, PG1211+143, Reeves et al. 2005). The nature of the plasma producing these features is currently unclear. Dadina et al. (2005) suggested that they
may be due to clumps of matter ejected from the disc, as expected in  the ``aborted jets" model proposed by Ghisellini et al. (2004). A similar scenario has been suggested by Turner et al. (2002, 2004) to explain shifted narrow {\it emission} lines in Seyfert spectra. The detailed geometry of the flow is as yet very obscure, but the transitory behaviour of most of the spectral features  seems to 
be in favour of an unsteady absorber rather than, say, a continuous wind. 
 
Whatever their origin, observations of redshifted Fe K absorption lines clearly have the potential
to add greatly to our knowledge of the gas flow in the innermost regions of AGN. 
  
\subsection{Conclusions from the models}
\label{subsec:discu_model_335}
Our simple inflow models lead us to two primary conclusions. 
Firstly, a large scale radial inflow which is highly ionised by a central 
source is not consistent with the data: although a strong inverse P Cygni
Fe~{\sc xxvi} K$\alpha$ line is predicted by such a model, it is too broad 
and insufficiently redshifted to explain the Mrk~335 observation.

Secondly, the absorption line can be well described by an inflow
which extends over only a limited radial extent, approximately 24 -- 48~$r_{g}$
from the black-hole. 
In this model, the red-shift of the absorption line is dominated by velocity
shift rather than gravitational shift and the
flow is highly ionised, explaining why it only
produces detectable absorption lines in the X-ray band. 
It is interesting to note that quasi-spherical flows with 
significantly higher densities (as
would be required to lower the ionisation state and obtain absorption lines
at softer energies) would be unphysical since the models considered here are
already close to the Eddington limit. Thus inflows which subtend a 
substantial fraction of the solid angle seen by the central black hole are only
ever likely to be readily detectable in the high energy Fe lines.

Clearly, the particular model and velocity-law 
considered -- that of spherical inflow from infinity -- is very
simplistic. 
More realistic scenarios which would result in rather similar 
absorption line profiles would include an infalling blob of gas or 
accretion column
occupying
approximately the same radial region as that considered in the successful
spherical model (i.e. $\approx$ 24 -- 48~$r_{g}$). However, more complete 
investigations of these scenarios 
would require more complex modelling and, much
more importantly, significantly higher quality data with greater spectral
resolution. 


\section*{Acknowledgments}
This paper is based on observations obtained with the {\it XMM-Newton} satellite,
an ESA science mission with contributions funded by ESA Member States and USA.
We thank the anonymous referee for his comments which contributed to improve this paper and we thank the 
{\it XMM-Newton} team for supporting the data analysis.
The authors wish to thank Prof. Andy Fabian for providing the relativistic 
blurring codes.
A.L.L. is grateful to Paul  O'Neill for many suggestions about the line significance analysis, to Giovanni Miniutti for stimulating discussion on this paper and to Maria Santos-Lleo for her help in the treatment of the data. 
This work was undertaken while S.A.S. was a UK Particle Physics and Astronomy 
Research Council supported postdoctoral research assistant at Imperial College London. 
A.L.L. acknowledges financial support  from the Royal Astronomical Society,  the Angelo Della Riccia Foundation and the Astrophysics Group at Imperial College London.

\end{document}